\documentclass[epsfig]{article}
\usepackage{amsfonts}

\usepackage{graphicx}
\usepackage{amsmath}

\input epsf.sty
\newtheorem{theorem}{Theorem}
\newtheorem{acknowledgement}[theorem]{Acknowledgement}

\input{tcilatex}
\makeatletter \@addtoreset{equation}{section}
\renewcommand{\theequation}{\thesection.\arabic{equation}}

\begin{document}

\title{\rightline{\mbox {\normalsize {Lab/UFR-HEP/0306/GNPHE/0307}}}\vspace{1cm}%
\textbf{Explicit Analysis of} \textbf{Kahler Deformations in }\\
\textbf{4D} $\mathcal{N}\mathbf{=}1$\textbf{\ Supersymmetric Quiver Theories}}
\author{Malika Ait Ben Haddou$^{1,2,3}$ and El Hassan Saidi$^{2,3}$\thanks{%
E-mail: h-saidi@fsr.ac.ma} \\
{\small 1-Unit\'{e} Alg\`{e}bre \& G\'{e}ometrie, D\'{e}partement de
Math\'{e}matique, Facult\'{e} des Sciences, M\`{e}knes, Morocco}\textit{,}\\
{\small 2-Lab/UFR-High Energy Physics, Physics Department, Faculty of
Science, Rabat, Kingdom of Morocco.}\\
{\small 3- Groupement National de Physique des Hautes Energies, Si\`{e}ge
Focal, Facult\'{e} des Sciences de Rabat, Morocco}}
\maketitle

\begin{abstract}
Starting from the $\mathcal{N}=2$ SYM$_{4}$ quiver theory living on wrapped $%
N_{i}D5$ branes around $S_{i}^{2}$ spheres of deformed ADE fibered
Calabi-Yau threefolds (CY3) and considering deformations using \textit{%
massive} vector multiplets, we explicitly build a new class of $\mathcal{N}%
=1 $\ quiver gauge theories. In these models, the quiver gauge group $%
\prod_{i}U\left( N_{i}\right) $ is spontaneously broken down to $%
\prod_{i}SU\left( N_{i}\right) $ and Kahler deformations are shown to be
given by the real part of the integral $\left( 2,1\right) $ form of CY3. We
also give the superfield correspondence between\ the $\mathcal{N}=1$\ quiver
gauge models derived here and those constructed in hep-th/0108120 using
complex deformations. Others aspects of these two dual $\mathcal{N}=1$
supersymmetric field theories are discussed.

\end{abstract}

\tableofcontents

\newpage \newpage

\section{Introduction}

\noindent Recently four dimension $\mathcal{N}=1$ supersymmetric quiver
gauge theories have been subject to an intensive interest \cite{r11}-\cite
{r13}. These theories, which are engineered in different but dual ways,
appear as low energy effective field theory of compactification of M-theory
on G2 manifolds and type II string compactification on threefolds preserving
1/8 of original supersymmetries \cite{r21}-\cite{r24}. A remarkable set of
such field theoretical system corresponds to those $4D$ $\mathcal{N}=1$
supersymmetric quiver gauge theories with gauge group $\prod_{i}U\left(
N_{i}\right) $ and which are obtained through deformations of $4D$ $\mathcal{%
N}=2$ $\prod_{i}U\left( N_{i}\right) $\ supersymmetric quiver gauge theories
living on $D5$\ branes wrapped on $ADE$ fibered Calabi-Yau threefolds (CY3)
\cite{r31,r32}. Two classes ( with and without monodromies) of such $4D$ $%
\mathcal{N}=1$ supersymmetric quiver gauge theories, following from the
complex deformation of $4D$ $\mathcal{N}=2$ supersymmetric quiver gauge
theories, has been constructed in \cite{r41,r42}. In this paper, we want to
derive their mirrors using Kahler deformations rather than complex ones.
Note that from the geometric point of view, this kind of dual models follow
naturally using algebraic geometry methods and mirror symmetry exchanging
Kahler and complex deformations; but from the supersymmetric field theory
view the situation is far from obvious and needs a careful treatment. We
will show, amongst others, that Kahler deformations in supersymmetric quiver
field theories require massive gauge prepotentials; that is a spontaneously
broken gauge symmetry $\prod_{i}U\left( N_{i}\right) $ down to $%
\prod_{i}SU\left( N_{i}\right) $ with all the features that go with this
behaviour and too particularly the implementation of a Higgs superpotential
and so adding further fundamental matters.

The presentation\ of this paper is as follows: In section 2, we describe the
$4D$ $\mathcal{N}=2$ $\prod_{i}U\left( N_{i}\right) $\ supersymmetric quiver
gauge theories living on $D5$\ branes wrapped on $ADE$ fibered Calabi-Yau
threefolds (CY3). We focus our attention on the special example of $U\left(
N\right) $ gauge theory engineered on a A$_{1}$ fibered CY3 and use a
simplest path involving the minimal degrees of freedom. Extension to ADE
geometries is straightforward and some of its aspect may be found in \cite
{r51}. In section 3, we develop the study of the $4D$ $\mathcal{N}=1$ $%
\prod_{i}U\left( N_{i}\right) $\ supersymmetric quiver gauge theories
following from complex deformations of the $\mathcal{N}=2$ SYM$_{4}$ quiver
models. In section 4, we consider the mirror of the previous $\mathcal{N}=1$
supersymmetric quiver gauge by using Kahler deformations rather than complex
ones. In section 5, we give our conclusion. Note that we will work in $%
\mathcal{N}=1$ superspace and make use of both real superspace $\left(
x,\theta ,\overline{\theta }\right) $ techniques as well as chiral ones $%
\left( x\pm i\theta \sigma \overline{\theta },\theta ,\overline{\theta }%
\right) $. For technical details; see for instance \cite{r51,r52}.

\section{4D $\mathcal{N}=2$ SYM$_{4}$ quiver theories: A$_{1}$ model}

The $\mathcal{N}=2$ supersymmetric A$_{1}$ quiver theory in four dimensions
involves the following $\mathcal{N}=1$ degrees of freedom: (\textbf{i}) A $%
U\left( N\right) $ gauge multiplet $V$ which we take in the WZ gauge as $%
V=-\theta \sigma ^{\mu }\overline{\theta }A_{\mu }-i\overline{\theta }%
^{2}\theta \lambda +i\theta ^{2}\overline{\theta }\overline{\lambda }+\frac{1%
}{2}\theta ^{2}\overline{\theta }^{2}D$. This superfield has the special
features
\begin{equation}
V^{3}=0;\qquad V=\frac{1}{N}U\mathrm{Y}+\sum_{a=1}^{N^{2}-1}V_{a}T^{a};%
\qquad U=Tr\left( V\right)  \label{01}
\end{equation}
which will be needed later on. Here $\mathrm{Y}\sim \mathrm{I}_{id}$ is the
abelian U$\left( 1\right) $ generator of $U\left( N\right) $ and $\left\{
T^{a}\right\} $ refer to the $SU\left( N\right) $ traceless generators. (%
\textbf{ii}) A chiral multiplet $\Phi $ in the adjoint representation of the
gauge group $U\left( N\right) $. We will refer to it as adjoint matter and
has the two following decompositions.
\begin{equation}
\Phi =\frac{1}{N}\Theta \mathrm{Y}+\sum_{a=1}^{N^{2}-1}\Phi _{a}T^{a};\qquad
\Phi =\phi +\theta \psi +\theta ^{2}F;\qquad \overline{\mathcal{D}}\Phi =0,
\label{02}
\end{equation}
where $\mathcal{D}$ stands for the supersymmetric covariant derivative; $%
\left\{ \mathcal{D},\overline{\mathcal{D}}\right\} \sim 2\partial _{\mu }$,
and $\Theta =Tr\left( \Phi \right) $ is the U$\left( 1\right) $ part of the
adjoint of U$\left( N\right) $. We have also $\left[ \Theta \mathrm{Y},\Phi %
\right] =\left[ \Theta \mathrm{Y},U\mathrm{Y}\right] =\left[ \Theta \mathrm{Y%
},V\right] =0$. For late use, we will focus on the supersummetric vacuum
with a preserved SU$\left( N\right) $ gauge symmetry; that is matrix
superfields with vevs such as$\ <\Phi _{a}>=0$,\ but $<\Theta >\neq 0$. Note
that the computation of $Tr\left[ \Phi ^{m}\right] $ in terms of $\Theta $
and $\Phi _{a}$ involves SU$\left( N\right) $ Casimirs; however due to $%
<\Phi _{a}>=0$ the vev of $Tr\left[ \Phi ^{m}\right] $ simplifies to $Tr%
\left[ \left( \frac{1}{N}\Theta \mathrm{Y}\right) ^{m}\right] =N^{m-1}\Theta
^{m}$ and so superpotentials type $W\left( \Phi \right)
=\sum_{1}^{n+1}\delta _{m}Tr\Phi ^{m}$ reduces to a polynom in the U$\left(
1\right) $ superfield $\Theta $. (\textbf{iii}) Four chiral multiplets $%
Q_{\left( \pm ,\pm \right) }$ with the following $U\left( 1\right) \times
SU\left( N\right) $ charges: $Q_{\left( +,+\right) }\equiv Q_{+}$ and $%
Q_{\left( -,+\right) }\equiv P_{-}$ are in the representation $\left( \pm
1,N\right) $ and $Q_{\left( +,-\right) }\equiv P_{+}$ and $Q_{\left(
-,-\right) }\equiv Q_{-}$ are in the representation $\left( \pm 1,\overline{N%
}\right) $. The antichiral superfields are in the complex conjugate of these
representations. For convenience, we will work with the normalization of the
U$\left( 1\right) $ charge as $\left[ \mathrm{Y},Q_{\pm }\right] =2Q_{\pm }$
and $\left[ \mathrm{Y},P_{\pm }\right] =-2P_{\pm }$. These matter
superfields have, in the chiral basis, the following $\theta $-expansions,
\begin{equation}
Q_{\pm }=q_{\pm }+\theta \psi _{\pm }+\theta ^{2}f_{\pm };\qquad P_{\pm
}=p_{\pm }+\theta \eta _{\pm }+\theta ^{2}l_{\pm },  \label{03}
\end{equation}
where $q_{\pm }$, $p_{\pm }$ and so on stand for component fields. Note that
the chiral composites $Q_{+}Q_{-}$ and $P_{+}P_{-}$ are in the $U\left(
N\right) $ adjoint representation and may be expanded as in eqs(\ref{01},\ref
{02}). The same is valid for the hermitian composites $Q_{\pm }Q_{\pm
}^{\ast }$ and $P_{\pm }P_{\pm }^{\ast }$. Note also that these four $Q_{\pm
}$ and $P_{\pm }$\ chiral multiplets form two $\mathcal{N}=2$
hypermultiplets; one of them encodes the transverse coordinates of
D5-branes; it describes their positions in the ten dimensional type IIB
string space, and the other is the usual moduli associated with the Kahler
deformation of the A$_{1}$ singulariy $\cite{z0}$.

\textbf{Action}:\textbf{\qquad }The superspace lagrangian density $\mathcal{L%
}_{N=2}\left( A_{1}\right) $ describing the $\mathcal{N}=2$ dynamics of the
previous superfields reads as,

\begin{eqnarray}
\mathcal{L}_{N=2}\left( A_{1}\right) &=&\mathcal{L}_{g}\left( V\right)
-2\zeta \int d^{4}\theta U+\mathcal{L}_{ad}\left( \Phi \right) -\int
d^{2}\theta \left( \beta \Theta +Tr\left[ \Phi \left(
Q_{-}Q_{+}-P_{-}P_{+}\right) \right] +hc\right)  \notag \\
&&+\int d^{4}\theta Tr\left[ Q_{-}^{\ast }e^{-2V}Q_{-}+Q_{+}^{\ast
}e^{2V}Q_{+}+P_{-}^{\ast }e^{-2V}P_{-}+P_{+}^{\ast }e^{2V}P_{+}\right]
\label{04}
\end{eqnarray}
where $\mathcal{L}_{g}\left( V\right) $ and $\mathcal{L}_{ad}\left( \Phi
\right) $ are respectively the gauge covariant lagrangian densities for the $%
U\left( N\right) $ vector multiplets and adjoint matter superfields. The
coupling constants $\zeta $ and $\beta $ are respectively real and complex
parameters. They have both a field theoretical and geometric meanings and
will play a crucial role in the present study. The supersymmetric scalar
potential reads in terms of the auxiliary fields as $\mathcal{V}=\frac{1}{2}%
Tr\left( D^{2}\right) +Tr\left( FF^{\ast }\right) +Tr\left( f_{\pm }^{\ast
}f_{\pm }\right) +Tr\left( l_{\pm }^{\ast }l_{\pm }\right) $ and the moduli
space of its vacuum configuration is given by the following eqs,
\begin{equation}
\zeta =r_{1}-r_{2};\qquad \beta =t_{1}-t_{2}  \label{05}
\end{equation}
where we have set
\begin{eqnarray}
r_{1} &=&<Tr\left( Q_{+}Q_{+}^{\ast }+P_{+}P_{+}^{\ast }\right) >;\qquad
t_{1}=<Tr\left( Q_{-}Q_{+}\right) >  \notag \\
r_{2} &=&<Tr\left( Q_{-}Q_{-}^{\ast }+P_{-}P_{-}^{\ast }\right) >;\qquad
t_{2}=<Tr\left( P_{-}P_{+}\right) >.  \label{06}
\end{eqnarray}
These parameters have a geometric interpretation in terms of Kahler and
complex moduli of the A$_{1}$ fiber of the CY3. The real parameter $\zeta $
is the volume of the blown up sphere and the complex constant\ is just the
so called holomorphic volume of the complex deformation of A$_{1}$. In
algebraic geometry, this means,

\begin{equation}
\zeta =\int_{S_{r}^{2}}\mathcal{J}^{\left( 1,1\right) };\qquad \beta
=\int_{S_{h}^{2}}\mathcal{\omega }^{\left( 2,0\right) };\qquad \beta ^{\ast
}=\int_{S_{a}^{2}}\mathcal{\omega }^{\left( 0,2\right) },  \label{07}
\end{equation}
where $\mathcal{J}^{\left( 1,1\right) }$ and $\mathcal{\omega }^{\left(
2,0\right) }$ are respectively the Kahler and complex holomorphic forms on
the A$_{1}$ surface. Note in passing that the algebraic geometry eq of the
complex deformed of the A$_{1}$ fiber of the CY3 reads as,
\begin{equation}
x^{2}+y^{2}+\left( z-\Delta t\right) \left( z+\Delta t\right) =0,
\label{070}
\end{equation}
where $\Delta t$ stands for the holomorphic volume $\left(
t_{1}-t_{2}\right) $ of the complex deformation which, by help of eq(\ref{05}%
), is also equal to $\beta $ and so eq(\ref{070}) may be rewritten as $%
x^{2}+y^{2}+z^{2}=\beta ^{2}$.

\begin{figure}[tbp]
\begin{center}
\epsfxsize=7cm\epsffile{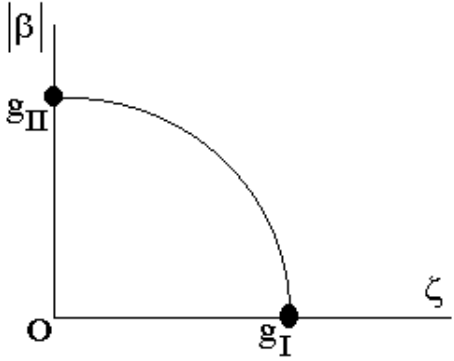}
\end{center}
\caption{\textit{In this figure, we represent the projection of a flow }%
\textrm{g}$_{N=2}=\mathrm{g}_{N=2}\left( \protect\vartheta \right) $\textit{%
\ on the }$\left( \protect\zeta ,\left| \protect\beta \right| \right) $%
\textit{\ plane.\ The black dot on the }$\protect\zeta $\textit{\ axis
represents }\textrm{g}$_{N=2}^{\left( I\right) }$\textit{\ and the one on
the }$\left| \protect\beta \right| $\textit{\ axis represents }\textrm{g}$%
_{N=2}^{\left( II\right) }$\textit{.}}
\label{ma}
\end{figure}

\textbf{Mirror N=2 models}:\qquad On the supersymmetric field theory side,
the $\zeta $ and $\left| \beta \right| $\ parameters are involved in the $%
\mathcal{N}=2$ SYM gauge coupling constant \textrm{g}$_{N=2}^{\left(
SYM\right) }\equiv $\textrm{g}$_{N=2}$ which read, in terms of the type IIB
string coupling $\mathrm{g}_{s}$, as,
\begin{equation}
\mathrm{g}_{N=2}=\sqrt{\frac{\mathrm{g}_{s}}{\QTR{sl}{V}}};\qquad \QTR{sl}{V}%
=\sqrt{\zeta ^{2}+\beta \overline{\beta }},  \label{08}
\end{equation}
Note that from the above relation, one sees that the $\mathcal{N}=2$ SYM
coupling constant is a real two argument function; \textrm{g}$_{N=2}=\mathrm{%
g}_{N=2}\left( \zeta ,\left| \beta \right| \right) $, which we shall naively
rewrite as \textrm{g}$_{N=2}\left( \zeta ,\beta \right) $. Accordingly, one
may think about this gauge coupling constant as describing a flow of $%
\mathcal{N}=2$ SYM models interpolating between two extreme models I and II
with respective gauge coupling constants \textrm{g}$_{N=2}^{\left( I\right)
} $ and \textrm{g}$_{N=2}^{\left( II\right) }$. The first is,

\begin{equation}
\mathrm{g}_{N=2}^{\left( I\right) }=\sqrt{\frac{\mathrm{g}_{s}}{\QTR{sl}{V}%
_{I}}};\qquad \QTR{sl}{V}_{I}=\sqrt{\zeta ^{2}};\qquad \beta =0,  \label{09}
\end{equation}
with blown up volume $\QTR{sl}{V}_{I}$ and the second involves pure
holomorphic volume $\QTR{sl}{V}_{II}$ type Weil-Peterson as,
\begin{equation}
\mathrm{g}_{N=2}^{\left( II\right) }=\sqrt{\frac{\mathrm{g}_{s}}{\QTR{sl}{V}%
_{II}}};\qquad \QTR{sl}{V}_{II}=\sqrt{\beta \overline{\beta }};\qquad \zeta
=0.  \label{010}
\end{equation}
Setting $\zeta =\rho \cos \vartheta $\ and $\left| \beta \right| =\rho \sin
\vartheta $; with the spectral parameter $\vartheta $ bounded as$\ 0\leq
\vartheta \leq \frac{\pi }{2}$, one gets an explicit relation for this $%
\mathcal{N}=2$ gauge coupling constant flow $\mathrm{g}_{N=2}\mathrm{=g}%
_{N=2}\left( \vartheta \right) =\sqrt{\frac{\mathrm{g}_{s}}{\QTR{sl}{V}%
\left( \vartheta \right) }}$. In this view, the theories I and II with
respective gauge couplings \textrm{g}$_{N=2}^{\left( I\right) }$ and \textrm{%
g}$_{N=2}^{\left( II\right) }$ correspond to $\vartheta =0$ and $\frac{\pi }{%
2}$, they are mapped to each other under mirror symmetry acting as $%
\vartheta \rightarrow \frac{\pi }{4}-\vartheta $; see figure \ref{ma}.

In A$_{1}$ geometric language, the $\mathcal{N}=2$ gauge models I
corresponds to the blowing up of A$_{1}$ surface in CY3; but zero
holomorphic deformations, $\int_{S^{2}}\mathcal{\omega }^{\left( 2,0\right)
}=0$. The compact part of the A$_{1}$ singularity $x^{2}+y^{2}+z^{2}=0$ gets
a non zero volume as $\left( \func{Re}x\right) ^{2}+\left( \func{Re}y\right)
^{2}+\left( \func{Re}z\right) ^{2}=\zeta $. This positive Kahler parameter $%
\zeta $ is same as in the superfield action eq(\ref{04}). To fix the ideas $%
\zeta $ can be imagined of as corresponding to the derivative of a
special
Kahker deformation $K\left( h,\overline{h}\right) $ where $h$ and $\overline{%
h}$ are Higgs fields to be specified later on; see eq(\ref{24}). In other
words $\zeta =\partial K_{FI}/\partial C$ where $K_{FI}$ is a \textit{linear}
Kahler deformation as $K_{FI}\sim \zeta \left( C+U\right) $ and where $C=%
\frac{\upsilon ^{\ast }H+\upsilon \overline{H}}{\upsilon \upsilon ^{\ast }}$%
; see eqs(\ref{33}-\ref{36}). In the present paper, $\zeta $ should be
thought of as just the leading case of a \textit{non linear} Kahler
superpotential $K\left( H,\overline{H}\right) $ so that $\left( \func{Re}%
x\right) ^{2}+\left( \func{Re}y\right) ^{2}+\left( \func{Re}z\right)
^{2}=\zeta $ get replaced by $\left( \func{Re}x\right) ^{2}+\left( \func{Re}%
y\right) ^{2}+\left( \func{Re}z\right) ^{2}=K^{\prime }\left( h,\overline{h}%
\right) $. Along with this Kahler analysis, one may also consider its mirror
description using complex deformation of A$_{1}$ singularity. In this case
the resulting $\mathcal{N}=2$ gauge model II corresponds exactly to the
reverse of previous situation. Here $\int_{S^{2}}\mathcal{J}^{\left(
1,1\right) }=0$ but $\int_{S^{2}}\mathcal{\omega }^{\left( 2,0\right) }\neq 0
$. As before the A$_{1}$ singularity $x^{2}+y^{2}+z^{2}=0$ gets now a
holomorphic volume as $x^{2}+y^{2}+z^{2}=\beta ^{2}$ where $\beta $\ is as
in the super action eq(\ref{04}). Here also this $\beta $ appears as the
derivative of \textit{linear} complex deformation as $W_{FI}\sim \beta \phi $
which in general should be thought of as just the leading case of a \textit{%
non linear} polynomial superpotential $W\left( \phi \right) $ so that $%
x^{2}+y^{2}+z^{2}=\beta ^{2}$ extends to $x^{2}+y^{2}+z^{2}=W^{\prime
^{2}}\left( \phi \right) $ constituting a non trivial fibered deformed A$_{1}
$ in the CY3 we are interested in here. Note that $x^{2}+y^{2}+z^{2}=W^{%
\prime ^{2}}\left( \phi \right) $ describes a singular conifold in $\mathbb{C%
}^{4}$; no complex deformation of this conifold\footnote{%
Complex deformation of conifold singularity involves a desingularisation
moduli $\mu $ as $x^{2}+y^{2}+z^{2}=W^{\prime ^{2}}\left( \phi \right) +\mu $
which is required in geometric transition at large $N$ \cite{z1}. Here we
have only $W^{\prime ^{2}}\left( \phi \right) $ (deformed A$_{1}$ moduli);
but no $\mu =0$.} is made here and so it should'nt be confused with
geometric transition scenario of \cite{r41}. Note also that extension of $%
\zeta $ and $\beta $\ to non linear $K^{\prime }$\ and $W^{\prime }$
respectively break $\mathcal{N}=2$ supersymmetry down to $\mathcal{N}=1$.
From the field theoretical point of view, these two models correspond to
choosing the corresponding vevs eqs(\ref{05}-\ref{06}) such that $%
t_{1}\left( \phi \right) =t_{2}\left( \phi \right) $ and $r_{1}\left(
c\right) \neq r_{2}\left( c\right) $ and inversely $t_{1}\left( \phi \right)
\neq t_{2}\left( \phi \right) $ and $r_{1}\left( c\right) =r_{2}\left(
c\right) $. The two symmetric situations indicate the existence of two
mirror $\mathcal{N}=1$ supersymmetric A$_{1}$ quiver gauge theories I and II
with gauge couplings \textrm{g}$_{N=1}^{\left( I\right) }$ and \textrm{g}$%
_{N=1}^{\left( II\right) }$. Let us first study complex deformations of the
previous $\mathcal{N}=2$ theory by introducing chiral superpotentials W$%
\left( \Phi \right) $. Later, we consider how mirror Kahler superpotentials
may be implemented.

\section{$\mathcal{N}=1$ A$_{1}$ quiver gauge theory I}

This theory is obtained by performing complex deformations of the lagrangian
density $\mathcal{L}_{N=2}\left( A_{1}\right) $ eq(\ref{04}). This is
equivalent to introducing an extra chiral superpotential W in the adjoint
matter superfield and too particularly in the U$\left( 1\right) $ factor $%
\Theta $ of adjoint matter $\Phi $ eq(\ref{02}). In doing so, the lagrangian
density $\mathcal{L}_{N=2}\left( A_{1}\right) $ becomes,
\begin{equation}
\mathcal{L}_{N=1}^{\left( I\right) }\left( A_{1}\right) =\mathcal{L}%
_{N=2}\left( A_{1}\right) +\left( \int d^{2}\theta W\left( \Theta \right)
+hc\right) .  \label{11}
\end{equation}
In this relation, $\mathcal{N}=2$ supersymmetry is explicitly broken down to
$\mathcal{N}=1$ due the presence of the non linear super potential W$\left(
\Theta \right) $; but $U\left( N\right) $ gauge invariance is still
preserved. The superpotential $W\left( \Theta \right) $\ generating complex
deformations has two basic features which in fact are inter-related and play
an important role at the quantum level: (\textbf{i}) the holomorphic
property $\frac{\partial W\left( \Theta \right) }{\partial \Theta ^{\ast }}%
=0 $, which permit to benefit from the power of algebraic geometry and (%
\textbf{ii}) chirality
\begin{equation}
\int d^{4}\theta W\left( \Theta \right) =0  \label{111}
\end{equation}
allowing miraculous simplifications. Comparing the above lagrangian density (%
\ref{11}) with eq(\ref{04}), one learns that complex deformation by the
superpotential $W\left( \Theta \right) $ corresponds to promoting the
previous complex FI type linear term with complex coupling constant $\beta $%
, namely $\beta \int d^{2}\theta \left( \Theta \right) $, \ to a more
general chiral superfunction $\int d^{2}\theta W\left( \Theta \right) $. As
a consequence $W^{\prime }\left( \Theta \right) $ is no longer constant as
in general it is $\Theta $ dependent. It follows then that the constant $%
\beta $ of section 2 is now promoted to a U$\left( N\right) $ gauge
invariant function $\mathcal{P}\left( \phi \right) $ as,
\begin{equation}
\beta \rightarrow \mathcal{P}^{\prime }\left( \phi \right) =\beta -W^{\prime
}\left( \phi \right)  \label{12}
\end{equation}
where $W^{\prime }\left( \phi \right) =<W^{\prime }\left( \Theta \right) >$.
Moreover, putting the relation (\ref{12}) back into the expression of the
SYM gauge coupling $\mathrm{g}$, one gets the following running $\mathcal{N}%
=1$ gauge coupling constant,
\begin{equation}
\mathrm{g}_{N=1}\left( \phi \right) =\mathrm{g}_{N=2}\left( \zeta ,\beta
;\phi \right) ;\qquad \mathrm{g}_{N=1}^{\left( I\right) }\left( \phi \right)
=\mathrm{g}_{N=2}\left( \zeta ,\beta =0;\phi \right) ;\qquad \mathrm{g}%
_{N=1}^{\left( II\right) }\left( \phi \right) =\mathrm{g}_{N=2}\left( \zeta
=0,\beta ;\phi \right) .  \label{13}
\end{equation}
Note that $\mathcal{N}=2$ supersymmetry is recovered at the critical point $%
\phi _{0}$ of the superpotential; $W^{\prime }\left( \phi _{0}\right) =0$,
and so by expanding around this critical point, one may compute the
deviations of the $\mathcal{N}=1$ gauge coupling from the $\mathcal{N}=2$
value.
\begin{equation}
\mathrm{g}_{N=1}\left( \phi \right) =\mathrm{g}_{N=2}\left( \beta \right)
-\left( \phi W^{\prime \prime }\frac{\partial \mathrm{g}_{N=2}\left( \beta
\right) }{\partial \beta }+hc\right) +O\left( \phi ^{2}\right) ,  \label{130}
\end{equation}
where we have set $\phi _{0}=0$. The leading term of the $\phi $\ expansion
of $\mathrm{g}_{N=1}\left( \phi \right) $ is $\mathrm{g}_{N=2}\left( \beta
\right) $ and the next one depends on $W^{\prime \prime }$. For free
massless adjoint matter $\mathrm{g}_{N=1}\left( \phi \right) =\mathrm{g}%
_{N=2}\left( \beta \right) $ up to second order of $\phi $ expansion.
Setting $\beta =0$, one gets the variation of the\ coupling constant $%
\mathrm{g}_{N=1}^{\left( I\right) }$ around the value of the $\mathcal{N}=2$
one. Moreover as the real coupling constant $\zeta $ has been untouched by
the extension eq(\ref{11}), it follows then that the defining eqs of the
moduli space of this $\mathcal{N}=1$ supersymmetric quiver gauge theory
reads as,
\begin{equation}
\zeta =r_{1}-r_{2};\qquad \mathcal{P}^{\prime }\left( \phi \right) =\beta
-W^{\prime }\left( \phi \right) =t_{1}\left( \phi \right) -t_{2}\left( \phi
\right) .  \label{14}
\end{equation}
One of the special feature of this expression is that under complex
deformation, eq(\ref{070}) becomes
\begin{equation}
x^{2}+y^{2}+z^{2}=\left( \mathcal{P}^{\prime }\left( \phi \right) \right)
^{2},  \label{140}
\end{equation}
showing that the CY3 is indeed a complex deformed A$_{1}$ surface fibered on
the plane parameterized by the complex variable $\phi $. Furthermore using
the relation (\ref{14}) and comparing with eqs(\ref{07}), it is not
difficult to see that the superpotential of the adjoint matter considered
above is in fact linked to CY3 complex moduli space as follows,

\begin{equation}
W\left( \phi \right) =\beta \phi -\int_{S^{2}\times J}\Omega ,  \label{15}
\end{equation}
where $\Omega =\omega \wedge d\tau $ is a $\left( 3,0\right) $-form on CY3
realized by an A$_{1}$ fiber on the complex plane and where one recognizes
the FI terms $\beta \phi $. Such analysis extends straightforwardly to all
ADE fibered CY3; with both finite and affine ADE geometries. This aspect and
other feature will be exposed in $\cite{r51}$.

\section{$\mathcal{N}=1$ A$_{1}$ quiver gauge theory II}

Applying mirror symmetry ideas to the above $\mathcal{N}=1$ A$_{1}$ quiver
gauge theory I, one expect to be able to build its superfield theoretical
dual by starting from the the lagrangian density $\mathcal{L}_{N=2}\left(
A_{1}\right) $ eq(\ref{04}) and use Kahler deformations as,
\begin{equation}
\mathcal{L}_{N=1}^{\left( II\right) }\left( A_{1}\right) =\mathcal{L}%
_{N=2}\left( A_{1}\right) +\delta _{Kahler}\mathcal{L}.  \label{20}
\end{equation}
In superspace, $\delta _{Kahler}\mathcal{L}$ involve integration over the
full superspace measure and reads as,
\begin{equation}
\delta _{Kahler}\mathcal{L}=\int d^{4}\theta \mathcal{K},  \label{21}
\end{equation}
where $\mathcal{K}$ is a Kahler superpotential; that is some real
superfunction we still have to specify. In what follows, we show that $%
\mathcal{K}$ has much to do with massive gauge superfields.

\textbf{Massive gauge prepotential}:\qquad Although natural from geometric
point of view due to mirror mirror symmetry exchanging Kahler and complex
deformations of CY3 \cite{r61}, the superfield theoretical formulation of
the dual theory II is far from obvious. The point is that in the derivation
of $\mathcal{N}=1$ quiver gauge theories I, the promotion of $\beta $\ to
chiral superpotentials $W\left( \phi \right) $ uses the scalar moduli of
adjoint matter $\Theta $. However for the Kahler deformations we are
interested in here, one cannot use $\Theta $ by deforming the kinetic energy
density $\int d^{4}\theta \left( \Theta ^{\ast }\Theta \right) $ to,
\begin{equation}
\int d^{4}\theta \mathcal{K}_{adj}\left( \Theta ^{\ast }\Theta \right) ,
\label{22}
\end{equation}
where $\mathcal{K}_{adj}\left( \Theta ^{\ast }\Theta \right) $\ is a Kahler
superpotential for adjoint matter. A field theoretical reason for this is
that $\Theta $\ does not couple to the abelian U$\left( 1\right) $ gauge
prepotential of the $U\left( N\right) $\ gauge symmetry. The introduction of
Kahler deformations for the $Q_{\pm }$ and $P_{\pm }$ fundamental matters
as,
\begin{equation}
\int d^{4}\theta \mathcal{K}_{fund}\left( Q_{\pm }^{\ast }e^{\pm 2V}Q_{\pm
}+P_{\pm }^{\ast }e^{\pm 2V}P_{\pm }\right)
\end{equation}
does not solve the problem any more since this leads essentially to quite
similar relations to eqs(\ref{05}-\ref{07}). The adjunction of
superpotentials for fundamental matters does not work as well because it
breaks $SU\left( N\right) $ gauge symmetry down to subgroups and this is
ruled out by the A$_{1}$ fibered CY3 we are considering here. However there
is still an issue since a careful analysis for the Kahler analogue of the
chiral superpotential of complex deformations of theory I reveals that the
difficulty we encounter in theory II is not a technical one. It is linked to
the fact that in 4D $\mathcal{N}=1$ supersymmetric gauge theory II, the $%
\mathcal{N}=1$ massless gauge multiplet $\left( \frac{1}{2},1\right) $ has
no scalar moduli that could play the role of the coordinate of the complex
one dimension base of CY3. This is then a serious problem; but fortunately
not a basic one since it may be overcome by considering massive $\mathcal{N}%
=1$ gauge multiplets $U^{\left( mass\right) }$,
\begin{equation}
U^{\left( massive\right) }\sim \left( 0,\frac{1}{2}^{2},1\right) _{M},
\label{220}
\end{equation}
which have scalars contrary to massless gauge prepotentials. But how this
issue may be implemented in the original $\mathcal{N}=2$ supersymmetric
quiver gauge theory we started with? The answer is by spontaneously breaking
the abelian gauge sub-invariance as $U\left( N\right) \longrightarrow
SU\left( N\right) $. For general ADE geometries, the spontaneous breaking of
the quiver gauge symmetry should be as $\prod_{i}U\left( N_{i}\right) $ $%
\longrightarrow \prod_{i}SU\left( N_{i}\right) $. Using this result, one
still has to overcome the two following apparent difficulties.

\textbf{Two more things}:\qquad (\textbf{1}) From geometric point of view,
we know that the variable $\tau $ parameterizing the complex one dimension
base ( plane) of the CY3 is associated with the complex scalar modulus of
the adjoint matter multiplet $\Phi $ as shown on $\left( 0^{2},\frac{1}{2}%
\right) $,
\begin{equation}
\tau \qquad \leftrightarrow \qquad <Tr\Phi >=<\Theta >=\phi .  \label{23}
\end{equation}
In the case of $\mathcal{N}=1$ massive gauge multiplets U$^{\left(
mass\right) }$, one has only one scalar modulus and it is legitimate to ask
from where does come the lacking scalar? This is a crucial question since
one needs one more scalar to be able to parameterize the two dimension base
of CY3. The answer to this question is natural in massive QFT$_{4}$; the
missing scalar degree is in fact hidden in the\ $\mathcal{N}=1$ on shell
massive gauge representation; it is just the longitudinal degree of freedom
of the massive spin one particle A$_{\mu }$. This a good point in the right
direction; but we still need to know how to extract this hidden scalar. The
right answer to this technical difficulty follows from a remarkable feature
of $\mathcal{N}=1$ supersymmetric theory which require complex manifolds
\cite{r52}. In the language of supersymmetric field theoretical
representations, the real scalar $c$ appearing in $\left( 0,\frac{1}{2}%
^{2},1\right) _{M}$ should in fact be thought of as the real part of a
complex field $h$ as $c\sim h+h^{\ast }$ where now $h$ is the scalar
component of chiral ( Higgs) superfield,
\begin{equation}
H=h+\theta \psi +\theta ^{2}F  \label{24}
\end{equation}
which one suspects justly to$\ $be the right modulus for parameterizing the
base of CY3. (\textbf{2}) The second thing concerns the way to implement the
massive vector multiplet into a $\mathcal{N}=2$ supersymmetric quiver gauge
theory we started with. The answer is to think about the $\mathcal{N}=1$
massive gauge multiplet $\left( 0,\frac{1}{2}^{2},1\right) _{M}$ as itself
following from the decomposition of a $\mathcal{N}=2$ massive gauge
multiplet $\left( 0^{5},\frac{1}{2}^{4},1\right) _{M}$ as shown on the
following decomposition,
\begin{equation}
\left( 0^{5},\frac{1}{2}^{4},1\right) _{M}\rightarrow \left( 0,\frac{1}{2}%
^{2},1\right) _{M}\oplus \left( 0^{2},\frac{1}{2}\right) \oplus \left( 0^{2},%
\frac{1}{2}\right) ,  \label{25}
\end{equation}
where $\left( 0^{2},\frac{1}{2}\right) \oplus \left( 0^{2},\frac{1}{2}%
\right) $ are two chiral multiplets. Moreover as the $\mathcal{N}=1$ massive
gauge multiplet $\left( 0,\frac{1}{2}^{2},1\right) _{M}$ may also be
decomposed as the sum of a $\mathcal{N}=1$ massless gauge multiplet and a $%
\mathcal{N}=1$ chiral superfield, one then end with the following spectrum:
(a) a massless abelian gauge prepotential $U$ and (b) three chiral
multiplets $H_{0,\pm }$ as shown here below,
\begin{equation}
\left( 0^{5},\frac{1}{2}^{4},1\right) _{M}\rightarrow \left( \frac{1}{2}%
,1\right) \oplus \left( 0^{2},\frac{1}{2}\right) _{+}\oplus \left( 0^{2},%
\frac{1}{2}\right) _{0}\oplus \left( 0^{2},\frac{1}{2}\right) _{-},
\label{26}
\end{equation}
where the charges $0,\pm 1$ appearing at the bottom of the matter multiplets
refer to charges under the abelian gauge factor of the U$\left( N\right) $
gauge symmetry.

\textbf{The N=1 quiver gauge Model II}:\qquad This supersymmetric model
involves the following $\mathcal{N}=1$ degrees of freedom: (\textbf{a}) A $%
U\left( N\right) $ gauge multiplet $V$ which has an abelian part $U$ as in
eqs(\ref{01}) and an SU$\left( N\right) $ part $V_{a}=Tr\left( T_{a}V\right)
$. (\textbf{b}) A chiral multiplet $\Phi $ in the adjoint representation of
the gauge group $U\left( N\right) $. The abelian part $\Theta $ of this
adjoint matter is identified with the neutral superfield appearing in the
decomposition (\ref{26}). The non abelian term is given by the set $\Phi
_{a}=Tr\left( T_{a}\Phi \right) $. (\textbf{c}) Four chiral matter
superfields $Q_{\pm }$ and $P_{\pm }$ transforming in the fundamental
representations of the $U\left( 1\right) \times SU\left( N\right) $ gauge
symmetry as in eqs(\ref{03}). All these superfields exist already in the
original $\mathcal{N}=2$ model we have described in section 2. (\textbf{d})
Two more chiral multiplets $H_{\pm }$ carrying \ $\pm 2$ charges under the
abelian symmetry of the gauge group and transform as scalars with respect to
$SU\left( N\right) $. The $H_{\pm }$ superfields are associated with the
multiplets $\left( 0^{2},\frac{1}{2}\right) _{\pm }$ appearing in the
decomposition eq(\ref{26}). In summary, we have the following $\mathcal{N}=1$
superfield spectrum: (\textbf{i}) the quartet
\begin{equation}
U,\quad H_{0}\equiv \Theta ,\quad H_{+},\quad H_{-},  \label{27}
\end{equation}
which describe the degrees of freedom abelian massive $\mathcal{N}=2$
multiplet $\left( 0^{5},\frac{1}{2}^{4},1\right) _{M}$ eq(\ref{26}). The
chiral multiplets should be thought of as Higgs superfields and whose Kahler
superpotential,
\begin{equation}
\int d^{4}\theta \mathcal{K}_{Higgs}\left( H_{+}^{\ast
}e^{2U}H_{+}+H_{-}^{\ast }e^{-2U}H_{-}\right)
\end{equation}
is exactly what we need. (\textbf{ii}) The $SU\left( N\right) $ massless $%
\mathcal{N}=2$ vector multiplet which in terms of the $\mathcal{N}=1$
superfield language we are using here reads as $V_{a}$ and $\Phi _{a}$; and (%
\textbf{iii}) finally the two $\mathcal{N}=2$ hypermultiplets $Q_{\pm }$ and
$P_{\pm }$ describing fundamental matters. From this supersymmetric
representation analysis, one learns that dynamics of massive $\mathcal{N}=2$
vector multiplet may be formulated in $\mathcal{N}=1$ superspace by starting
with a massless vector multiplet $U$ and three chiral ones $H_{0,\pm }$ as
introduced before. To get\ a massive gauge superfield, one gives non trivial
vevs to $H_{\pm }$; a fact which is achieved by introducing a superpotential
$\mathcal{W}_{ext}\left( H_{+},H_{0,},H_{-}\right) $ describing couplings
between chiral superfields. Since we are interested by the engineering of $%
\mathcal{N}=1$ quiver gauge theory using Kahler deformations, we will not
insist on having $\mathcal{N}=2$ supersymmetric couplings for Higgs
superfields. So we restrict the extra superpotential to $\mathcal{W}_{ext}=%
\mathcal{W}_{ext}\left( H_{+},H_{-}\right) $ with the two following
requirements: ($\alpha $) The full scalar potential $\mathcal{V}$ of the
supersymmertic gauge abelian model namely $\mathcal{V}=\frac{1}{2}%
D_{U}^{2}+F_{+}F_{+}^{\ast }+F_{0}F_{0}^{\ast }+F_{-}F_{-}^{\ast }$ vanishes
in the vacuum ( $D_{U}=F_{0,\pm }=0$) and ($\beta $) at least one of the
chiral superfield $H_{\pm }$ acquires a vev when minimising $\mathcal{V}$ ($%
\frac{\partial \mathcal{V}}{\partial h_{\pm }}=0$). Let take these vevs as,
\begin{equation}
<H_{+}>=\upsilon ;\qquad <H_{-}>=0,  \label{h1}
\end{equation}
where $\upsilon $ is a complex parameter. A simple candidate for gauge
invariant Higgs superpotential fulfilling features ($\alpha $) and ($\beta $%
) is $\mathcal{W}_{ext}=mH_{+}H_{-}$ with mass $m$ linked to $\zeta $ and $%
\upsilon $; i.e $m=m\left( \zeta ,\upsilon \right) $. With this in mind one
can go ahead to work out the Kahler deformation program. In what follows, we
describe the main lines and omit details.

\textbf{The Action for N=1 quiver Theory II}:\qquad From the above
discussion, it follows that the lagrangian density $\mathcal{L}%
_{N=1}^{\left( II\right) }\left( A_{1}\right) =\mathcal{L}_{N=2}\left(
A_{1}\right) +\delta _{Kahler}\mathcal{L}$ eq(\ref{20}) of the $\mathcal{N}%
=1 $ supersymmetric quiver model II is given by the following
superfunctional,
\begin{equation}
\mathcal{L}_{N=1}^{\left( II\right) }\left( A_{1}\right) =\mathcal{L}%
_{N=2}\left( A_{1}\right) -\left( \int d^{2}\theta \mathcal{W}_{ext}\left(
H_{+},H_{-}\right) +hc\right) +\int d^{4}\theta \left[ \mathcal{K}\left(
H_{+}^{\ast }e^{U}H_{+}\right) +H_{-}^{\ast }e^{-U}H_{-}\right] ,  \label{32}
\end{equation}
In this relation we have endowed the matter superfield $H_{+}$ with a Kahler
potential $\mathcal{K}\left( H_{+}^{\ast }e^{U}H_{+}\right) $ and left $%
H_{-} $ with a flat geometry. The introduction of a Kahler potential for $%
H_{-}$ does add nothing new since it is $H_{+}$ that is eaten by the gauge
prepotential after symmetry breaking. $\mathcal{K}\left( H_{+}^{\ast
}e^{U}H_{+}\right) $ is then crucial in the derivation of $\mathcal{N}=1$
quiver theories II; it is the mirror of W$\left( \Phi \right) $ of $\mathcal{%
N}=1$ quiver gauge theories I.

\section{More Results}

In the lagrangian density $\mathcal{L}_{N=2}\left( A_{1}\right) $ eq(\ref{04}%
) of the $\mathcal{N}=2$ quiver theory, Kahler deformations are encoded in $%
2\zeta \int d^{4}\theta \left( U\right) $. This term should appear as a
particular Kahler deformation in the $\mathcal{N}=1$ supersymmetric quiver
theory II encoded in the term $\int d^{4}\theta $ $\mathcal{K}\left(
H_{+}^{\ast }e^{U}H_{+}\right) $. Choosing $\mathcal{K}$ as follows,
\begin{equation}
\mathcal{K}_{FI}\left( H_{+}^{\ast }e^{U}H_{+}\right) =2\zeta \ln \left(
H_{+}^{\ast }e^{U}H_{+}\right) ,  \label{33}
\end{equation}
one recovers FI deformation; thanks to chirality $\int d^{4}\theta \left(
H_{+}\right) =0$. Therefore Kahler deformations $\mathcal{R}$ that are
mirror to the chiral potential $\mathcal{P}\left( \mathrm{\Theta }\right)
=\beta \Phi -W\left( \mathrm{\Theta }\right) $\ we have used in eq(\ref{12})
read in general as,
\begin{equation}
\mathcal{R}\left( \mathrm{Y}\right) =2\zeta \ln \left( \mathrm{Y}\right) -%
\mathcal{K}\left( \mathrm{Y}\right)  \label{34}
\end{equation}
where $\mathrm{Y}=H_{+}^{\ast }e^{U}H_{+}$. In this result similarity
between Kahler and complex deformation is perfect. It is a consequence of
mirror symmetry in this super QFT and may also be rederived from the
analysis of the lagrangian density (\ref{32}). The appearance of this
composite hermitian superfield $\mathrm{Y}$ is not fortuitous; it is just a
manifestation of the massive gauge prepotential we have discussed before.
Indeed parameterizing $H_{+}$ as,
\begin{equation}
H_{+}=\upsilon \exp \left( \frac{H}{\upsilon }\right) ,  \label{35}
\end{equation}
where now $H$ describe quantum fluctuation, we have for $\mathrm{Y}$,
\begin{equation}
\mathrm{Y}=\upsilon \upsilon ^{\ast }\exp \left( \frac{\upsilon ^{\ast
}H+\upsilon H^{\ast }}{\upsilon \upsilon ^{\ast }}+U\right) .  \label{36}
\end{equation}
But the term $\frac{\upsilon ^{\ast }H+\upsilon H^{\ast }}{\upsilon \upsilon
^{\ast }}+U$ in the exponential is nothing but the massive gauge
prepotential U$^{\left( mass\right) }$ of eq(\ref{220}). Eq(\ref{36}) and (%
\ref{34}) give actually the relation between massive gauge
multiplet and Kahler deformations. Moreover the defining eqs for
the moduli space of the supersymmetric vacua of Kahler
deformations in $\mathcal{N}=1$ quiver theories II following from
(\ref{32}) reads as,
\begin{equation}
\mathcal{R}^{\prime }\left( c\right) =\left( \mathrm{r}_{1}\mathrm{-r}%
_{2}\right)  \label{37}
\end{equation}
where $c=\frac{\upsilon ^{\ast }h+\upsilon h^{\ast }}{\upsilon \upsilon
^{\ast }}$ and where $\mathcal{R}\left( c\right) =\mathcal{R}\left[ \mathrm{y%
}\left( c\right) \right] $ and $\mathrm{y}\left( c\right) =\upsilon \upsilon
^{\ast }\exp \left( c\right) $ are as follows,
\begin{equation}
\mathcal{R}\left( \mathrm{y}\right) =\left[ \zeta \ln \mathrm{y}-\mathcal{K}%
\left( \mathrm{y}\right) \right] ;\qquad \mathrm{y}\left( c\right) =\left[
\upsilon \exp \left( \frac{h}{\upsilon }\right) \right] \left[ \upsilon
^{\ast }\exp \left( \frac{h^{\ast }}{\upsilon ^{\ast }}\right) \right]
\overline{=}w\overline{w}.  \label{38}
\end{equation}
Eq(\ref{37}) shows that the blown sphere depends on the coordinate of the
base of CY3. Like before, $\mathcal{N}=2$ supersymmetry is explicitly broken
down to $\mathcal{N}=1$ except at the critical point $c_{0}$ of $\mathcal{R}%
\left( c\right) $ where it is recovered; but $U\left( N\right) $ gauge
invariance is spontaneously broken down to $SU\left( N\right) $. In terms of
the quantum fluctuation superfields $H$ and $H^{\ast }$ eq(\ref{35}), the
critical point $\mathcal{R}^{\prime }\left( C_{0}\right) =0$ is translated
to,
\begin{equation}
\left( \overline{\upsilon }\frac{\partial \mathcal{R}\left(
H_{0},H_{0}^{\ast }\right) }{\partial H_{0}}+\upsilon \frac{\partial
\mathcal{R}\left( H_{0},H_{0}^{\ast }\right) }{\partial H_{0}^{\ast }}%
\right) =0  \label{39}
\end{equation}
This relation should be thought of as the analogue of $\frac{\partial W}{%
\partial \Phi ^{\ast }}=0$ in complex deformations. One can also compute the
variation of the $\mathcal{N}=1$ runing gauge coupling $\mathrm{g}%
_{N=1}\left( c\right) =\mathrm{g}_{N=1}\left( \zeta ,\beta ;c\right) $
around the value of the $\mathcal{N}=2$ one $\mathrm{g}_{N=2}\left(
c_{0}\right) $ living at the critical point $\mathcal{K}^{\prime }\left(
c_{0}\right) =0$. One finds, for a generic point on the $\mathcal{N}=2$
supersymmetric flow $\mathrm{g}=\mathrm{g}\left( \vartheta \right) $, the
following dual formula to eq(\ref{130}),
\begin{equation}
\mathrm{g}_{N=1}\left[ c\right] =\mathrm{g}_{N=2}\left( \zeta ,\beta \right)
-\left( c-c_{0}\right) \mathcal{K}^{\prime \prime }\left( \mathrm{c}%
_{0}\right) \frac{\partial \mathrm{g}_{N=1}\left( \zeta \right) }{\partial
c_{0}}+O\left[ \left( c-c_{0}\right) ^{2}\right] .  \label{40}
\end{equation}
Note\ by the way that one may also work out the mirror of eqs(\ref{140},\ref
{15}). Spliting $x$, $y$ and $z$ as $x=x_{1}+ix_{2}$ and so on, one may
decompose the complex surface $x^{2}+y^{2}+z^{2}=0$ into a compact part $%
x_{1}^{2}+y_{1}^{2}+z_{1}^{2}=0$ and a non compact one. Deformations of
compact part as $x_{1}^{2}+y_{1}^{2}+\left( z_{1}-\Delta \mathrm{r}\right)
\left( z_{1}+\Delta \mathrm{r}\right) =0$ and substituting $\Delta \mathrm{r}
$ is as in eqs(\ref{37}), one gets the real analogue of eq(\ref{15}) namely
\begin{equation}
x^{2}+y^{2}+z^{2}=\left( \mathcal{R}^{\prime }\left( c\right) \right) ^{2}.
\label{42}
\end{equation}
Geometrically, this means that $\mathcal{R}\left( c\right) $ generates
Kahler deformations of the CY3 and one can check that $\mathcal{R}\left(
c\right) $ is given by the following,
\begin{equation}
\mathcal{R}\left( c\right) =\zeta c+\int_{S^{2}\times J}\mathcal{K}^{\left(
2,1\right) }+\int_{S^{2}\times \overline{J}}\mathcal{K}^{\left( 1,2\right) },
\label{43}
\end{equation}
where $\mathcal{K}^{\left( 2,1\right) }$ and $\mathcal{K}^{\left( 1,2\right)
}$ are respectively $\left( 2,1\right) $ and $\left( 1,2\right) $\ forms on
CY3 and where one recognizes the usual FI term $\zeta c$ of the $\mathcal{N}%
=1$ abelian gauge theories. The correspondence between the two theories is
then perfect.

\section{Conclusion}

In this paper, we have developed the field theoretic analysis of
deformations of $4D$ $\mathcal{N}=2$ quiver gauge theories living in $D5$\
branes wrapped on $A_{1}$ fibered CY3. Though it looks natural by using
algebraic geometry methods and mirror symmetry exchanging complex and Kahler
moduli, such study is far from obvious on the field theoretical side. After
noting that the gauge coupling constant \textrm{g}$_{N=2}$ of such a theory
is given by a spectral flow
\begin{equation}
\mathrm{g}_{N=2}=\mathrm{g}_{N=2}\left( \vartheta \right) ;\qquad \tan
\vartheta =\frac{\left| \beta \right| }{\zeta }\qquad 0\leq \vartheta \leq
\frac{\pi }{2},
\end{equation}
with $\mathrm{g}_{N=2}\left( 0\right) $ and $\mathrm{g}_{N=2}\left( \frac{%
\pi }{2}\right) $\ respectively associated with pure Kahler and pure complex
deformations in the $A_{1}$\ fiber, we have considered deformations in the
full moduli space of CY3. For complex deformations, geometry implies that we
have the two following: (a) If deformations are restricted to the ADE
fibers, then $\mathcal{N}=2$ supersymmetry is preserved, up to a global
shift of energy and (b) If they cover the full CY3, then $\mathcal{N}=2$
supersymmetry is broken down to $\mathcal{N}=1$. Mirror symmetry implies
that similar results are also valid for Kahler deformations. On the
superfield theoretical view, this corresponds to adding appropriate
superpotential ( complex and Kahler) terms in the original $\mathcal{N}=2$
SYM$_{4}$. We have studied complex deformations of $\mathcal{N}=2$
supersymmetric quiver theories by using the method of \cite{r31} and given
amongst others the field expansion of the $\mathcal{N}=1$ running gauge
coupling constant \textrm{g}$_{N=1}$ around \textrm{g}$_{N=2}$. We have also
developed the explicit analysis for Kahler deformations of $\mathcal{N}=2$
supersymmetric quiver theories and shown that such real deformations require
massive gauge prepotentials $U^{\left( mass\right) }$ implying in turn a
spontaneously broken U$\left( N\right) $ gauge symmetry down to SU$\left(
N\right) $. We have worked out this program explicitly and shown amongst
others that Kahler deformations are given by the following.
\begin{equation}
\delta _{Kahler}\mathcal{L}_{\mathcal{N}=2}=\int d^{4}\theta \mathcal{R}%
\left( U^{\left( mass\right) }\right) ,  \label{44}
\end{equation}
where $\mathcal{R}\left( U^{\left( mass\right) }\right) $ is as in eqs(\ref
{38}).\ This relation, which generalize naturally for all ADE fibered CY3,
should be compared with the usual complex deformation involving the chiral
superpotential of adjoint matter,
\begin{equation}
\delta _{\func{complex}}\mathcal{L}_{\mathcal{N}=2}=\int d^{4}\theta
\mathcal{P}\left( \Phi \right) .  \label{45}
\end{equation}
with $\mathcal{P}\left( \Phi \right) $ as in eq(\ref{12}). The
analysis we have developed in this paper has the remarkable
property of being explicit. It allows superfield realizations of
geometric properties of CY3 and offers a powerful method to deal
with 4D $\mathcal{N}=1$ supersymmetric field theories living on
wrapped D5. Through this explicit field theoretic study, one also
learn that, on the $\mathcal{N}=1$ supersymmetric field
theoretical side, mirror symmetry acts by exchanging the roles of
adjoint matters $\Phi $ and massive gauge prepotentials $U^{\left(
mass\right) }$. On the geometric side, we have shown that Kahler
deformations, generated by the real superfield $\mathcal{R}\left(
U^{\left( mass\right) }\right) $, are given by the real part of
the integral of a $\left( 2,1\right) $ form on CY3 as shown on
eq(\ref{43}). This analysis may be also extended to incorporate D3
branes by considering affine ADE symmetries. Details on aspects of
this study as well as other issues may be found in \cite{r51}.

\begin{acknowledgement}
We thank Protars III/ CNRST/Rabat, for support and A. Belhaj for
discussions.
\end{acknowledgement}

\end{document}